\documentclass[
superscriptaddress,
preprint,
 amsmath,amssymb,
 aps,
]{revtex4-1}

\usepackage{graphicx}
\usepackage{dcolumn}
\usepackage{bm}

\begin{document}


\title{Unfaulting mechanisms of Frank loops in fluorite oxides}
\author{Miaomiao Jin}
\email{Corresponding author: M. Jin (mmjin@psu.edu)}
\affiliation{Department of Nuclear Engineering, The Pennsylvania State University, 218 Hallowell Building, University Park, 16802 PA, USA}
\author{Jilang Miao}
\affiliation{Department of Nuclear Engineering, The Pennsylvania State University, 218 Hallowell Building, University Park, 16802 PA, USA}
\author{Yongfeng Zhang}%
\affiliation{Department of Engineering Physics, University of Wisconsin-Madison, 1500 Engineering Dr, Madison, WI 53706, USA}
\author{Marat Khafizov}%
\affiliation{Department of Mechanical and Aerospace Engineering, The Ohio State University, 201 W 19th Ave, Columbus, OH 43210, USA}
\author{Kaustubh K. Bawane}%
\affiliation{Idaho National Laboratory, 2525 Fremont Ave, Idaho Falls, ID 83402, USA}
\author{Boopathy Kombaiah}%
\affiliation{Idaho National Laboratory, 2525 Fremont Ave, Idaho Falls, ID 83402, USA}
\author{David H. Hurley}%
\affiliation{Idaho National Laboratory, 2525 Fremont Ave, Idaho Falls, ID 83402, USA}%

\begin{abstract}
Unfaulting of Frank loops in irradiated fluoride oxides are of significance to microstructural evolution. However, the mechanisms have not been directly observed. To this end, we utilize molecular dynamics to reveal the atomistic details related to the unfaulting process of interstitial Frank loop in ThO$_2$, which involve nucleation of single or multiple Shockley partial pairs at the loop circumference. The unfaulting is achieved via a synchronous shear of the partial pairs to remove the extrinsic stacking fault in the cation sublattice and the intrinsic stacking fault in the anion sublattice. The strong oxygen motion at the dislocation core may reduce the activation barriers of dislocation nucleation and migration. These findings provide a fundamental understanding of the transformation of faulted loops in irradiated ThO$_2$, and could be transferable to other fluorite systems.  

\begin{description}
\item[Keywords]
Dislocation loop, unfaulting, fluorite structure 
\item[Declaration]
No conflict of interest
\end{description}

\end{abstract}

\pacs{Valid PACS appear here}
\maketitle

Aggregation of radiation-induced defects can form dislocation loops, which are commonly recognized under transmission electron microscopes (TEM) in irradiated crystalline specimens \cite{chen2020effects,kubota2005transition}. In fluorite-structured compounds, such as UO$_2$, ThO$_2$, CeO$_2$, and ZrO$_2$, the character of dislocation loops identified in irradiated samples depends on the irradiation conditions. In UO$_2$ irradiated with heavy ions, only perfect loops ($\mathbf{b}=a_0/2\langle110\rangle$, where $a_0$ is the lattice parameter) are reported, while neutron irradiation leads to both perfect and faulted loops ($\mathbf{b}=1/3\langle111\rangle)$ loops. In Kr-irradiated ThO$_2$, both perfect and faulted loops are observed \cite{he2022dislocation}. In Kr- and Xe-irradiated CeO$_2$, only faulted loops are confirmed \cite{chen2013characterization}. In ZrO$_2$, faulted loops were identified under high-voltage electron microscope \cite{baufeld1993situ}. These loops are recognized as interstitial loops, and the formation is believed to be mainly due to the diffusion and clustering of radiation-induced cation defects \cite{chen2013characterization,he2022dislocation}. In some other cases with electron irradiation, faulted loops consisting of only oxygen interstitials were also proposed in CeO$_2$ \cite{yasunaga2008electron}, however, such a high unbalance of local charge may not be energetically stable \cite{ryazanov2002growth}. 

Elastic theory indicates that the perfect loop is more energetically favorable than the faulted loop at large sizes due to the energy penalty from the stacking fault \cite{hull2001introduction}, causing faulted loops thermodynamically unstable as size increases. As the perfect loop resultant from loop unfaulting is mobile to interact with other defects (e.g., dislocation networks, grain boundaries, voids, etc), the unfaulting process is of significance to understanding the microstructure evolution. In ZrO$_2$ at 200 $^o$C, faulted loops show fast growth and instantaneously transform into a perfect loop under high-voltage electron microscope \cite{baufeld1993situ}. In Kr-irradiated ThO$_2$, the transformation from faulted loops to perfect loops is implied in the loop statistics: i) loop size tends to saturate and ii) the loop density tends to decrease with increasing dose. Although various indirect results suggest the unfaulting, to the authors' knowledge, the unfaulting process in fluorite oxides has not yet been directly resolved in experiments and simulations. 

 As the microscopic details remain largely challenging in experiments, molecular dynamics (MD) simulations have been frequently used to examine the loop unfaulting process in various lattice structures \cite{kadoyoshi2007molecular,chen2021comprehensive,dai2022atomistic,chen2022combined}. Based on the understanding of face-centered cubic (FCC) metals, which are pertinent to fluorite structures, to remove the stacking fault enclosed by the loop, there are various mechanisms involving moving Shockley partials ($\mathbf{b} = a/6<112>$) to sweep the fault \cite{chen2022combined}. Removal of intrinsic stacking fault (SF) can be realized via a single sweep of Shockley partial, while for extrinsic SF, double sweeps of Shockley partials are required. The Shockley partials may nucleate at the loop line or within the loop. The resultant dislocation structure depends on the loop habit plane, stacking fault energy, and external stress \cite{chen2022combined}. Although the understanding of FCC systems is largely established, one key issue to resolve is whether such understanding is transferable to the fluorite oxides, where oxygen atoms may mediate the unfaulting process. 

To pursue a mechanistic understanding, we utilize large-scale MD simulations to probe the details of loop unfaulting in ThO$_2$, which is expected to apply to other fluorite systems. Here we use ThO$_2$ as a demonstration to alleviate the concern that unpaired $f$ electrons as in UO$_2$ and CeO$_2$ may cause serious inaccuracy in the interatomic potential with the loop structures. The classical MD method as implemented in LAMMPS \cite{plimpton2007lammps} is used with an embedded-atom method potential for ThO$_2$ by Cooper et al. \cite{cooper2014many}. This potential takes a fixed non-formal charge model for both Th (+2.2208) and O (-1.1104) atoms, and can well reproduce mechanical properties and defect energies as compared to first-principles calculations and experimental measurements. Periodic boundary conditions, and Nose-Hoover thermostat \cite{nose1984unified,hoover1985canonical} (if applicable) are utilized. All the constructed systems are charge neutral to avoid numerical issues resulting from periodic boundary conditions. 

\begin{figure}[!ht]
	\centering
	\includegraphics[width=0.7\textwidth]{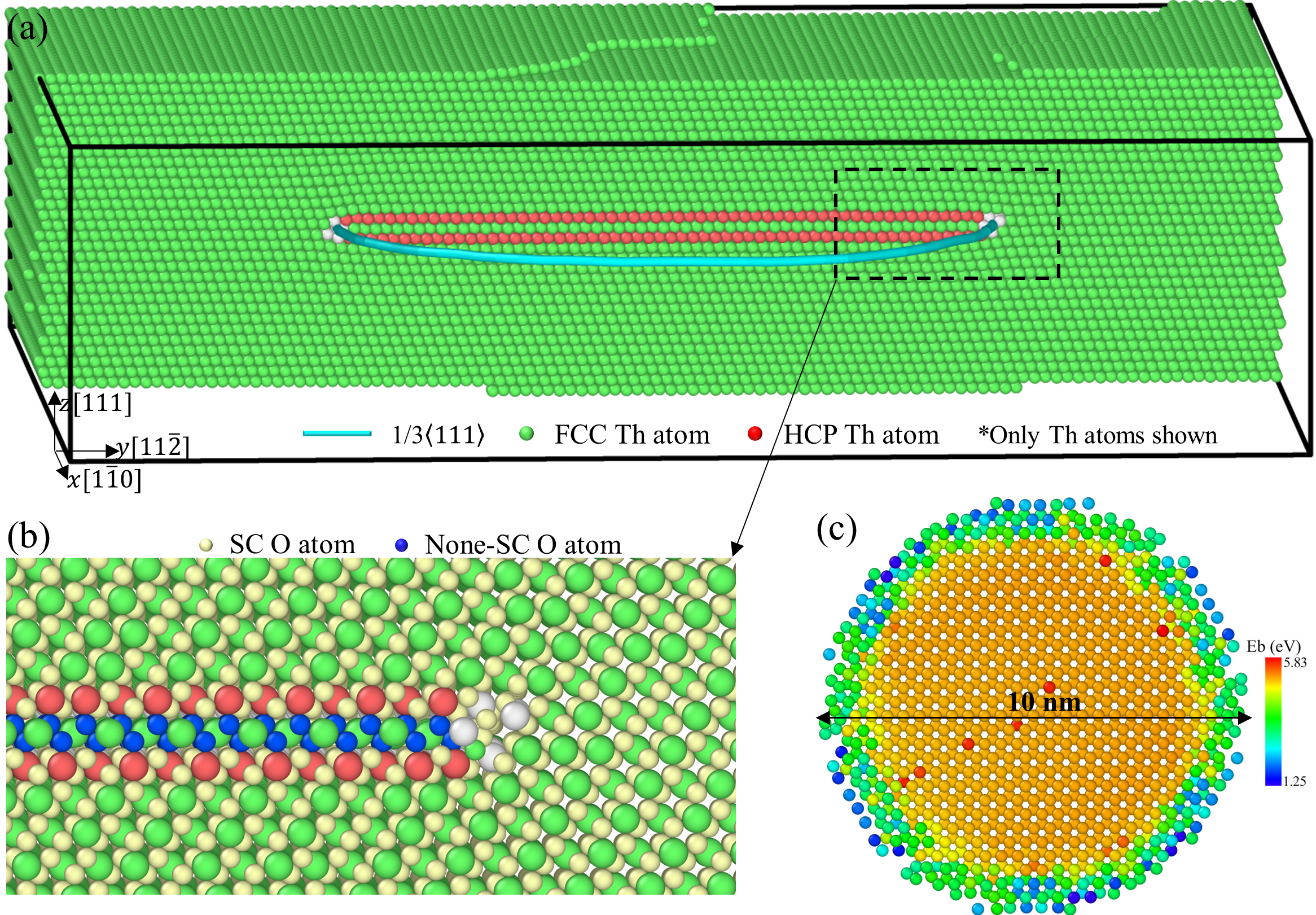}
	\caption{(a) Faulted dislocation loop configuration in ThO$_2$. (b) Enlarged local atom arrangement, showing the local stacking of Th and O layers. (c) Binding energies of individual oxygen atoms in the loop (loop Th atoms are not shown).}
	\label{fig:energies}
\end{figure}

To be consistent with neutron/ion-irradiated experiments, we only focus on loops of interstitial nature. To enable a strong driving force to initiate the loop transformation within the MD timescale, a large faulted loop (20 nm in diameter) is created, with a total of 974,685 atoms and a box size of around 40\,nm $\times$ 40\,nm $\times$ 10\,nm (FIG. \ref{fig:energies}a). This loop size guarantees a higher formation energy than that of a perfect loop, and such size is scarcely observed in ThO$_2$ \cite{he2022dislocation}. In fact, transformation may start to occur in nanometer-sized loops, as suggested in UO$_2$ \cite{chartier2016early,le2016empirical}. The loop is created by inserting three layers, i.e., O-Th-O {111} layers; such stacking has been suggested to be energetically favorable in other fluorite oxides, e.g., UO$_2$ \cite{le2016empirical,aidhy2011comparison}, CeO$_2$ \cite{aidhy2011comparison,chen2013characterization} and ZrO$_2$ \cite{baufeld1993situ}. The stacking of layers is carefully controlled to ensure correct stacking (FIG. \ref{fig:energies}b). Although hexagonal-shaped loops are frequently observed in FCC metals, the loops identified in fluoride oxides commonly appear in circular shape \cite{baufeld1993situ,he2013situ}, which is hence assumed in the current study. Atom visualization is accomplished with the OVITO package \cite{stukowski2009visualization}. 
 
In the current calculations, loop stoichiometry (i.e., each inserted layer has the exact same number of atoms to enforce loop charge neutrality) is imposed to avoid numerical issues. Such an assumption is partly supported by the MD simulations of loop formation in CeO$_2$ \cite{aidhy2011comparison}. However, the diffusivity of oxygen interstitials is much higher than that of Th interstitials \cite{colbourn1983calculated,he2022dislocation} and the formation of an initial loop may involve reconfiguration of small interstitial clusters \cite{jiang2022unraveling}. Hence, the dislocation loop can be off-stochiometric. To quantify how strongly these oxygen atoms bind to the loop, FIG. \ref{fig:energies}c shows a heatmap of the individual oxygen binding energy in a 10 nm diameter faulted loop by calculating the energy difference of the loop with respect to placing the loop constituent oxygen atom in the far field. The core oxygen atoms are much more strongly bonded than the periphery oxygen atoms due to the local lattice distortion at the dislocation core. As strong pipe diffusion was identified in UO$_2$ \cite{murphy2014pipe}, it is possible that under irradiation conditions, the loop tends to be oxygen-rich via trapping oxygen interstitials. This may reduce the energy barrier for loop unfaulting due to instability from charged loops, which is implied in the electron irradiation experiments of ion-irradiated ZrO$_2$, where fast loop growth and disappearance were observed when only oxygen defects are created \cite{ryazanov2002growth,yasuda2003radiation}.  

\begin{figure}[!ht]
	\centering
	\includegraphics[width=0.6\textwidth]{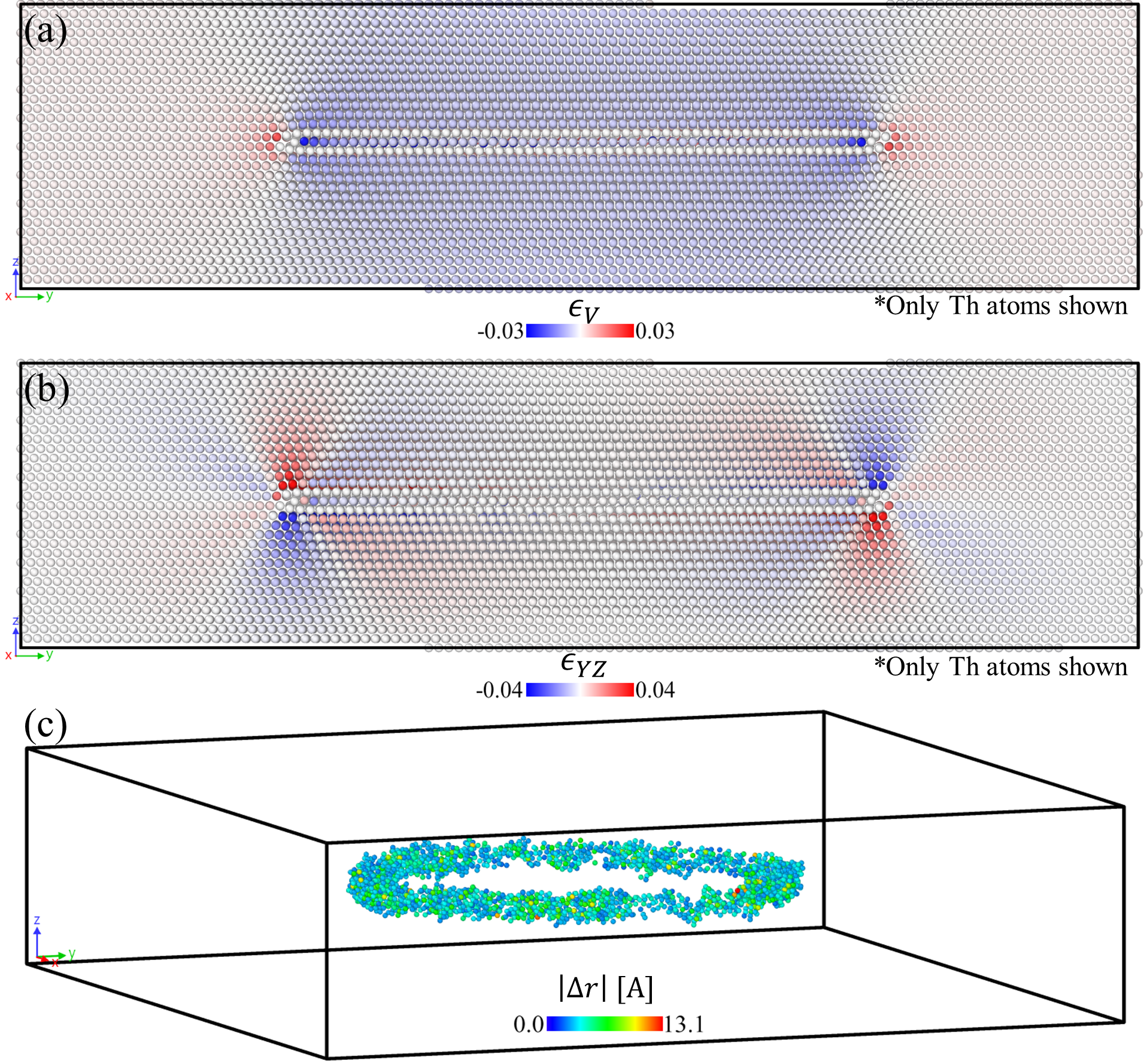}
	\caption{Distribution of volumetric strain $\epsilon_V$ (a), shear strain $\epsilon_{YZ}$ (b). (c) shows the oxygen displacement $\vert\Delta r\vert$ after 400 ps relaxation at 1500 K.}
	\label{fig:strain}
\end{figure}

The elastic strain field \cite{stukowski2012elastic} is computed near the loop as shown in FIG. \ref{fig:strain}a-b for volumetric strain ($\epsilon_V$) and shear strain ($\epsilon_{YZ}$), respectively. As expected, the strain is exceptionally high near the core region. The varying field of shear strain parallel to the plane indicates a possible double-lobe contrast under TEM, as similarly indicated in reference \cite{ryazanov2002growth} with diminished shear strain near the center. To reveal the effect of strain field on atom kinetics, system annealing at 1500 K and zero external pressure is performed. FIG. \ref{fig:strain}c demonstrates a heatmap of oxygen displacements (Th atoms are removed due to negligible displacement) after 400 ps. Large displacement of oxygen atoms mainly occurs at the loop line, consistent with the previous understanding of pipe diffusion in ThO$_2$ and UO$_2$ \cite{murch1987oxygen,murphy2014pipe}. Given the unstable state of oxygen atoms at the dislocation core, it is expected that Shockley partials are easier to nucleate at the loop circumference than within the loop.  

\begin{figure}[!ht]
	\centering
	\includegraphics[width=0.95\textwidth]{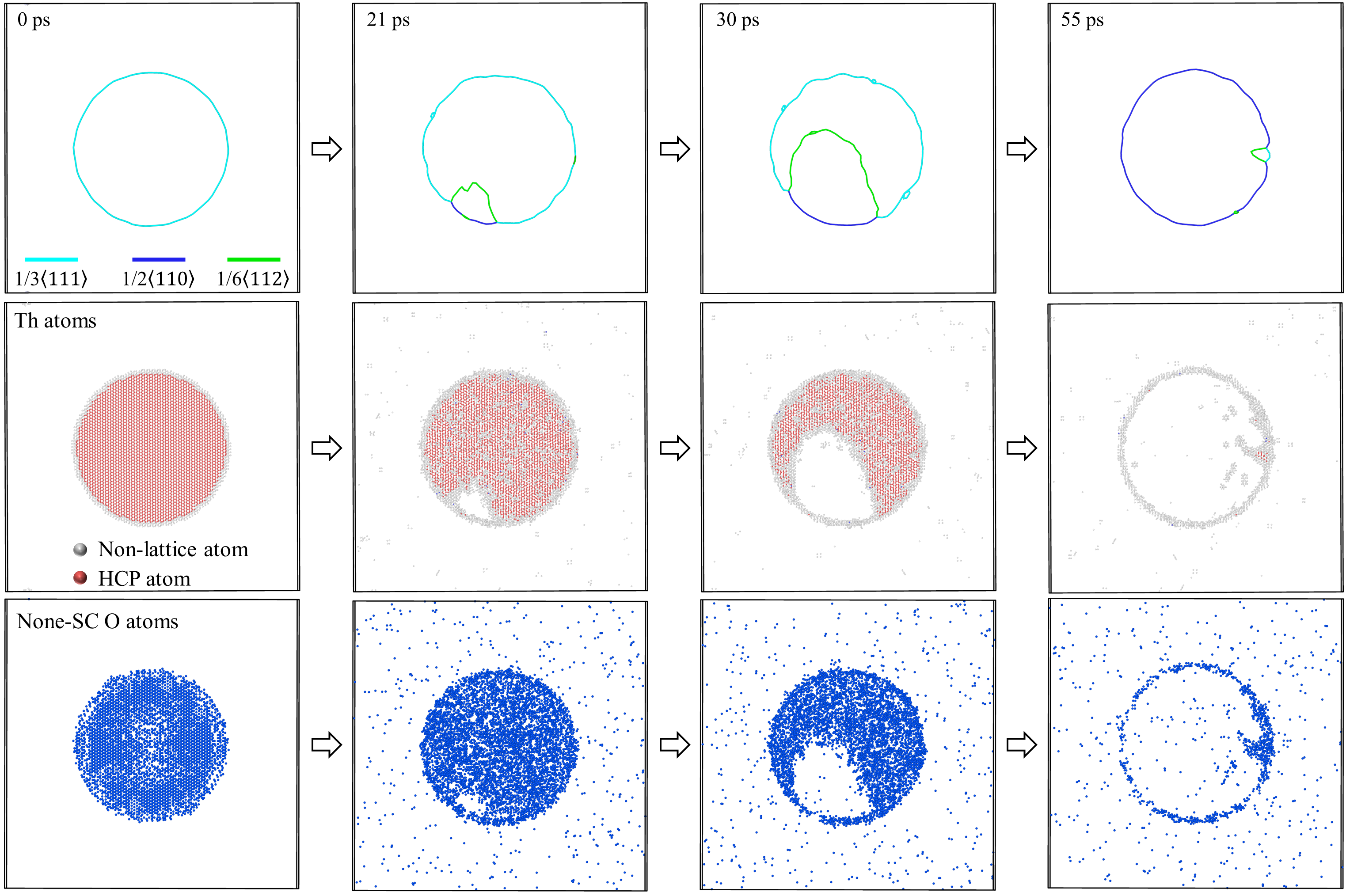}
	\caption{Loop unfaulting process with single Shockley partial nucleation site. Top row: dislocation types, middle row: Th stacking fault, bottom row: O stacking fault.}
	\label{fig:single_shockley}
\end{figure}

\begin{figure}[!ht]
	\centering
	\includegraphics[width=0.95\textwidth]{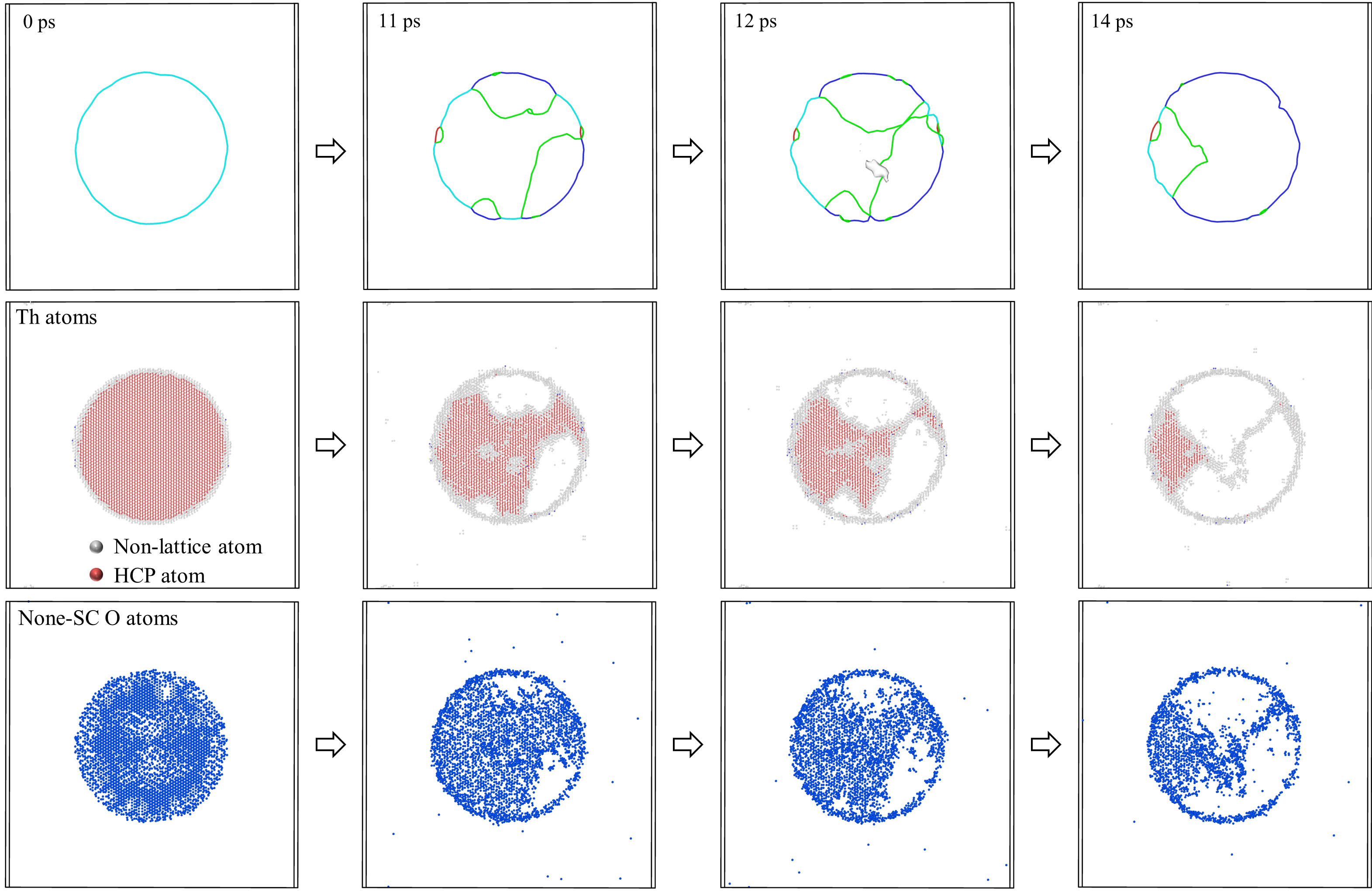}
	\caption{Loop unfaulting process with multiple Shockley partial nucleation sites. Top row: dislocation types, middle row: Th stacking fault, bottom row: O stacking fault.}
	\label{fig:multiple_shockley}
\end{figure}

The unfaulting process can be in principle initiated due to shear stress or thermal vibration \cite{kadoyoshi2007molecular,chen2021comprehensive}. Hence, two methods have been attempted to initiate the transformation, i.e., shear and high temperature. The shear is achieved by static shear of the simulation box, via iterative i) rigid 0.1 $\text{\AA}$ displacement of the top layer along $y$-axis with the bottom layer fixed, and ii) subsequent energy minimization. A sudden unfaulting due to mechanical instability occurs when the simulation box is sheared to 8.5 \% as a result of lattice translation, without any indication of Shockley partial nucleation (see Supplementary video 1). This phenomenon is consistent with previous simulations of loop unfaulting in FCC Cu \cite{kadoyoshi2007molecular}. It suggests that thermal activation is necessary to initiate the unfaulting process. Hence, the subsequent simulations are run at 1,500\,K at zero external pressure to capture the unfaulting process. For a reasonable computational cost, the initial configuration is slightly sheared (2.68 \%) along the $y$-axis, as this effectively reduces the energy barrier for the nucleation of Shockley partial at a specific temperature. For reference, high-temperature annealing up to 2,500 K of non-sheared structure does not exhibit the unfaulting process on the nanosecond timescale. FIG. \ref{fig:single_shockley} demonstrates the complete unfaulting process within a span of 55 ps: Shockley partial is nucleated from the loop line, and quickly sweeps across the fault plane loop, which ultimately converts the original loop to a perfect one (see Supplementary video 2). In another case with a 5.15 \% sheared simulation cell, a more complex unfaulting process is revealed, which involves the nucleation of multiple Shockley partials along the loop line (FIG. \ref{fig:multiple_shockley}). The propagation of the arc segments contributes to the unfaulting simultaneously, leading to reactions between Shockley partials within the loop plane. Since the nucleated Shockley partials are not necessarily of the same type within the (111) plane, the reaction can cause defect formation from local lattice incompatibility (see Supplementary video 3). At the end of the unfaulting, the defects gradually disappear due to atom diffusion. In both unfaulting patterns, by analyzing the faults in both the Th and O layers, two common points are identified: i) the extrinsic SF is removed in a single sweep of Shockley partial; and ii) Both Th and O sublattice stacking faults are removed in a synchronous manner. 


We note some important observations of loop unfaulting behavior in this system. First, it is feasible to have both single and multiple partial nucleation. The large-sized loop utilized in the current work provides a strong driving force to unfaulting. Such nucleation behavior is ultimately determined by the local stress and temperature-activated atomic events. It has also been reported in the unfaulting of both vacancy and interstitial faulted loops in Cu, Al, and Ni \cite{kubota2005transition,chen2021comprehensive,chen2022combined}. Second, unlike the commonly proposed mechanisms for loop unfaulting where a Shockley partial loop can be nucleated within the loop to explain experimentally observed loop contrast \cite{chen2021comprehensive,chen2022combined}, only Shockley partial nucleation at the loop perimeter is observed in this system; this is presumably due to that the strong strain field and oxygen mobility near the dislocation core (FIG. \ref{fig:energies}) can facilitate the partial nucleation upon thermal fluctuation. Third, we have not observed two-stage unfaulting as reported in metals \cite{kubota2005transition,kadoyoshi2007molecular,chen2022combined}, where a first set of Shockley partials sweeps the extrinsic stacking fault, converting it into an intrinsic stacking fault and a second set removes the remaining fault. To explain the non-existence of any intrinsic stacking fault ribbon, it is only possible that a joint Shockley partial pair sweep the fault simultaneously, with one shearing the above and the other shearing the below atoms, respectively. Hence, the total reaction is written as,
\begin{align*}
  1/6[\bar{1}2\bar{1}] & + 1/6[2\bar{1}\bar{1}] + 1/3[111] \rightarrow 1/2[110] \\
  A\delta & + B\delta + D\delta \rightarrow DC  ~ (\text{Thompson tetrahedron notation})
\end{align*}
Herein, $A\delta + B\delta \rightarrow \delta C (1/6[11\bar{2}])$, in the direction of applied shear. The obtuse angle between the two partials suggests attractive interaction, which can justify the close range of two joint partials. Based on energetic estimation proportional to $\mathbf{b}^2$, $\vert A\delta\vert ^2 + \vert B\delta\vert ^2 > \vert \delta C\vert ^2$, but overall $\vert A\delta \vert ^2 + \vert B\delta \vert ^2 + \vert D\delta \vert ^2 > \vert DC\vert ^2$, leading to energy reduction. 

\begin{figure}[!ht]
	\centering
	\includegraphics[width=0.75\textwidth]{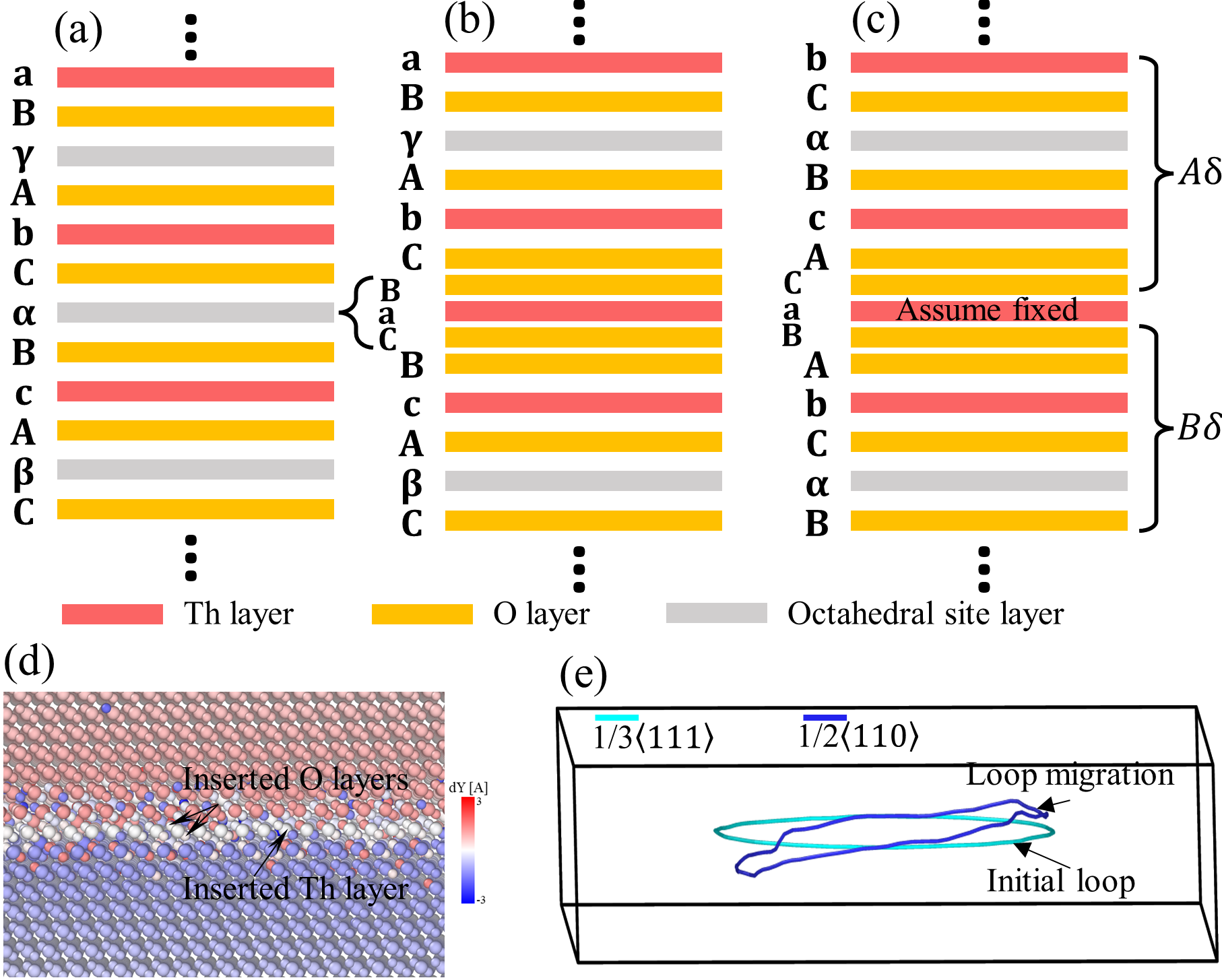}
	\caption{Stacking sequences for perfect lattice (a), with faulted loop (b), and with unfaulted loop (c). (d) shows a heatmap of displacement along $y$-axis ($dY$) after loop unfaulting. (e) shows loop migration over $\sim$45 ps after loop unfaulting. }
	\label{fig:mechanism}
\end{figure}

To explain the concurrent removal of faults in both Th and O layers given the sweep of a joint Shockley partial pair, the details of layer stacking are presented in the following. The layer stacking sequence along [111] is depicted in FIG. \ref{fig:mechanism}a for the perfect fluorite structure, where cation, anion, and octahedral site layers are interleaved to form an FCC stacking sequence \cite{le2016empirical}. We denote the sequence of cations, ...abc..., the anions ...ABC..., and the octahedral sites, ...$\alpha\beta\gamma$..., and the combined sequence is ...aB$\gamma$AbC$\alpha$BcA$\beta$C.... To construct the faulted loop, three layers (O-Th-O) are inserted into the octahedral site layer, forming ...aB$\gamma$AbC\textit{/BaC/}BcA$\beta$C... (FIG. \ref{fig:mechanism}b), where the italic font indicates the fault layers. Hence, the interstitial loop means extrinsic stacking fault for cation layers and intrinsic stacking fault for anion layers in the corresponding sublattices. Using the Thompson tetrahedron notation, on $(111)$ plane, only the following Shockley partials can be accommodated, i.e., $A\delta$, $\delta A$, $B\delta$, $\delta B$, $C\delta$, and $\delta C$. To align with simulation observations, with respect to the fault cation layer $a$, layers above and below $a$ must be simultaneously sheared by a Shockley partial, $1/6[\bar{1}2\bar{1}] $ $(A\delta)$ and $-1/6[2\bar{1}\bar{1}]$ $(-B\delta$, note the sign comes from the inverse view for the above and below), respectively, resulting in ...bC$\alpha$BcACaBAbC$\alpha$B... (FIG. \ref{fig:mechanism}c). Therefore, with respect to the layer $a$, the relative displacement of the adjacent layers (both cation and anion layers) should be around $a_0/\sqrt{6}$, and diminishes as distance increases; this is confirmed by the atom displacement field after unfaulting (FIG. \ref{fig:mechanism}d shows the displacement along $y$-axis). 

The O layers are generally commensurate with Th layers in displacement, nevertheless, large displacement for O atoms close to the whole loop plane is also observed (not shown in FIG. \ref{fig:mechanism}d). It is attributed to the pipe diffusion mechanism as the Shockley partial sweeps across the plane. Similar to dislocation nucleation, the high O diffusivity may reduce the energy barrier of dislocation migration as it can induce local off-stoichiometry, causing a metastable state of dislocation segments. After loop unfaulting, the perfect loop shows high mobility along the Burgers' vector direction (FIG. \ref{fig:mechanism}e) given the applied shear strain. Under irradiation conditions, these perfect loops can react with other extended defects to become dislocation networks. Importantly, defective cation and anion atoms are found only exist at the dislocation core of the perfect dislocation loops. Hence, unfaulting can be of significance to controlling the total defect accumulation in ThO$_2$. Similar to FCC metals, a thermodynamic critical size of faulted loops should exist in fluoride oxides, on the order of a few nanometers based on the formation energy of loops \cite{he2022dislocation}. Such critical size is generally smaller than that of metals due to its high stacking fault energy (e.g., 1.76 J/m$^2$ in ThO$_2$ \cite{he2022dislocation}, 1.808 J/m$^2$ in UO$_2$ \cite{le2016empirical}, and 78.5 mJ/m$^2$ in Cu \cite{heino1999stacking}). However, due to the nucleation barrier of Shockley partials and interaction with other radiation-induced defects, this thermodynamic critical size may not be particularly meaningful. For example, in heavy ion irradiated CeO$_2$, no perfect loop was observed for loops up to 20-30 nm in diameter, and in ThO$_2$ and UO$_2$ with various irradiation types, faulted loops are frequently observed beyond a few nanometer sizes \cite{he2022dislocation,whapham1965radiation,he2013situ,onofri2017full}. Finally, it is worth noting that the unfaulting, specifically partial nucleation, can be strongly affected by the radiation condition. For example, only faulted loops are identified in proton-irradiated ThO$_2$ \cite{bawane2021tem} while both types of loops are found in Kr-irradiated ThO$_2$ \cite{he2022dislocation}; this contrast could suggest that pronounced damage cascade disturbance assists in the partial nucleation, which warrants further investigation. 

In summary, with large-size stoichiometric circular Frank loops, we have identified two unfaulting patterns involving single and multiple Shockley partial pairs under combined shear strain and high temperature to explain the Frank loop unfaulting in fluorite ThO$_2$. Each Shockley pair is jointly moving in a synchronous manner to transform the stacking faults into perfect stacking. Cation and anion layers are commensurate during loop unfaulting and perfect loop migration. The high mobility of anions along dislocation line likely reduces the activation barriers of dislocation activities, which remains to be confirmed.  
\section*{Acknowledgments}
This work is supported by the Center for Thermal Energy Transport under Irradiation, an Energy Frontier Research Center funded by the U.S. Department of Energy, Office of Science, United States, Office of Basic Energy Sciences, and The Pennsylvania State University. 

\bibliographystyle{unsrt}
\bibliography{science}

\begin{thebibliography}{10}

\bibitem{chen2020effects}
Dongyue Chen, Kenta Murakami, Kenji Dohi, Kenji Nishida, Zhengcao Li, and Naoto
  Sekimura.
\newblock The effects of loop size on the unfaulting of frank loops in heavy
  ion irradiation.
\newblock {\em Journal of Nuclear Materials}, 529:151942, 2020.

\bibitem{kubota2005transition}
Alison Kubota and WG~Wolfer.
\newblock Transition pathways in the unfaulting of dislocation loops.
\newblock {\em Materials Science and Engineering: A}, 400:362--365, 2005.

\bibitem{he2022dislocation}
Lingfeng He, Tiankai Yao, Kaustubh Bawane, Miaomiao Jin, Chao Jiang, Xiang Liu,
  Wei-Ying Chen, J~Matthew Mann, David~H Hurley, Jian Gan, et~al.
\newblock Dislocation loop evolution in kr-irradiated tho2.
\newblock {\em Journal of the American Ceramic Society}, 105(8):5419--5435,
  2022.

\bibitem{chen2013characterization}
Wei-Ying Chen, Jianguo Wen, Marquis~A Kirk, Yinbin Miao, Bei Ye, Brian~R
  Kleinfeldt, Aaron~J Oaks, and James~F Stubbins.
\newblock Characterization of dislocation loops in ceo2 irradiated with high
  energy krypton and xenon.
\newblock {\em Philosophical Magazine}, 93(36):4569--4581, 2013.

\bibitem{baufeld1993situ}
Bernd Baufeld, Dietmar Baither, Ulrich Messerschmidt, Martin Bartsch, and
  Ingolf Merkel.
\newblock In situ study on the generation of radiation damage in cubic-zirconia
  in the high-voltage electron microscope.
\newblock {\em Journal of the American Ceramic Society}, 76(12):3163--3166,
  1993.

\bibitem{yasunaga2008electron}
K~Yasunaga, Kazuhiro Yasuda, Syo Matsumura, and T~Sonoda.
\newblock Electron energy-dependent formation of dislocation loops in ceo2.
\newblock {\em Nuclear Instruments and Methods in Physics Research Section B:
  Beam Interactions with Materials and Atoms}, 266(12-13):2877--2881, 2008.

\bibitem{ryazanov2002growth}
AI~Ryazanov, Kazuhiro Yasuda, C~Kinoshita, and AV~Klaptsov.
\newblock Growth and instability of charged dislocation loops under irradiation
  in ceramic materials.
\newblock {\em Journal of nuclear materials}, 307:918--923, 2002.

\bibitem{hull2001introduction}
Derek Hull and David~J Bacon.
\newblock {\em Introduction to dislocations}.
\newblock Butterworth-Heinemann, 2001.

\bibitem{kadoyoshi2007molecular}
Tomoko Kadoyoshi, Hideo Kaburaki, Futoshi Shimizu, Hajime Kimizuka, Shiro
  Jitsukawa, and Ju~Li.
\newblock Molecular dynamics study on the formation of stacking fault
  tetrahedra and unfaulting of frank loops in fcc metals.
\newblock {\em Acta materialia}, 55(9):3073--3080, 2007.

\bibitem{chen2021comprehensive}
Cheng Chen, Jing Zhang, and Jun Song.
\newblock Comprehensive study of vacancy frank loop unfaulting: atomistic
  simulations and predictive model.
\newblock {\em Acta Materialia}, 208:116745, 2021.

\bibitem{dai2022atomistic}
C~Dai, Q~Wang, P~Saidi, B~Langelier, CD~Judge, MR~Daymond, and MA~Mattucci.
\newblock Atomistic structure and thermal stability of dislocation loops,
  stacking fault tetrahedra, and voids in face-centered cubic fe.
\newblock {\em Journal of Nuclear Materials}, 563:153636, 2022.

\bibitem{chen2022combined}
Cheng Chen and Jun Song.
\newblock A combined atomistic-continuum study on the unfaulting of single and
  multi-layer interstitial dislocation loops in irradiated fcc and hcp metals.
\newblock {\em International Journal of Plasticity}, 152:103231, 2022.

\bibitem{plimpton2007lammps}
Steve Plimpton, Paul Crozier, and Aidan Thompson.
\newblock {LAMMPS}-large-scale atomic/molecular massively parallel simulator.
\newblock {\em Journal of Computational Physics}, 18, 2007.

\bibitem{cooper2014many}
MWD Cooper, MJD Rushton, and RW~Grimes.
\newblock A many-body potential approach to modelling the thermomechanical
  properties of actinide oxides.
\newblock {\em Journal of Physics: Condensed Matter}, 26(10):105401, 2014.

\bibitem{nose1984unified}
Shuichi Nos{\'e}.
\newblock A unified formulation of the constant temperature molecular dynamics
  methods.
\newblock {\em The Journal of Chemical Physics}, 81(1):511--519, 1984.

\bibitem{hoover1985canonical}
William~G Hoover.
\newblock Canonical dynamics: Equilibrium phase-space distributions.
\newblock {\em Physical Review A}, 31(3):1695, 1985.

\bibitem{chartier2016early}
A~Chartier, Claire Onofri, L~Van~Brutzel, Ch~Sabathier, O~Dorosh, and
  J~Jagielski.
\newblock Early stages of irradiation induced dislocations in urania.
\newblock {\em Applied Physics Letters}, 109(18):181902, 2016.

\bibitem{le2016empirical}
Arno Le~Prioux, Paul Fossati, Serge Maillard, Thomas Jourdan, and Philippe
  Maugis.
\newblock Empirical potential simulations of interstitial dislocation loops in
  uranium dioxide.
\newblock {\em Journal of Nuclear Materials}, 479:576--584, 2016.

\bibitem{aidhy2011comparison}
Dilpuneet~S Aidhy, Dieter Wolf, and Anter El-Azab.
\newblock Comparison of point-defect clustering in irradiated ceo2 and uo2: A
  unified view from molecular dynamics simulations and experiments.
\newblock {\em Scripta Materialia}, 65(10):867--870, 2011.

\bibitem{he2013situ}
Ling-Feng He, Mahima Gupta, Clarissa~A Yablinsky, Jian Gan, Marquis~A Kirk,
  Xian-Ming Bai, Janne Pakarinen, and Todd~R Allen.
\newblock In situ tem observation of dislocation evolution in kr-irradiated uo2
  single crystal.
\newblock {\em Journal of Nuclear Materials}, 443(1-3):71--77, 2013.

\bibitem{stukowski2009visualization}
Alexander Stukowski.
\newblock Visualization and analysis of atomistic simulation data with
  {OVITO}--the open visualization tool.
\newblock {\em Modelling and Simulation in Materials Science and Engineering},
  18(1):015012, 2009.

\bibitem{colbourn1983calculated}
EA~Colbourn and WC~Mackrodt.
\newblock The calculated defect structure of thoria.
\newblock {\em Journal of nuclear materials}, 118(1):50--59, 1983.

\bibitem{jiang2022unraveling}
Chao Jiang, Lingfeng He, Cody~A Dennett, Marat Khafizov, J~Matthew Mann, and
  David~H Hurley.
\newblock Unraveling small-scale defects in irradiated tho2 using kinetic monte
  carlo simulations.
\newblock {\em Scripta Materialia}, 214:114684, 2022.

\bibitem{murphy2014pipe}
Samuel~T Murphy, Eleanor~E Jay, and Robin~W Grimes.
\newblock Pipe diffusion at dislocations in uo2.
\newblock {\em Journal of Nuclear Materials}, 447(1-3):143--149, 2014.

\bibitem{yasuda2003radiation}
K~Yasuda, C~Kinoshita, S~Matsumura, and AI~Ryazanov.
\newblock Radiation-induced defect clusters in fully stabilized zirconia
  irradiated with ions and/or electrons.
\newblock {\em Journal of Nuclear Materials}, 319:74--80, 2003.

\bibitem{stukowski2012elastic}
Alexander Stukowski and A~Arsenlis.
\newblock On the elastic--plastic decomposition of crystal deformation at the
  atomic scale.
\newblock {\em Modelling and Simulation in Materials Science and Engineering},
  20(3):035012, 2012.

\bibitem{murch1987oxygen}
GE~Murch and C~Richard~A Catlow.
\newblock Oxygen diffusion in uo 2, tho 2 and puo 2. a review.
\newblock {\em Journal of the Chemical Society, Faraday Transactions 2:
  Molecular and Chemical Physics}, 83(7):1157--1169, 1987.

\bibitem{heino1999stacking}
Pekka Heino, L~Perondi, Kimmo Kaski, and E~Ristolainen.
\newblock Stacking-fault energy of copper from molecular-dynamics simulations.
\newblock {\em Physical Review B}, 60(21):14625, 1999.

\bibitem{whapham1965radiation}
AD~Whapham and BE~Sheldon.
\newblock Radiation damage in uranium dioxide.
\newblock {\em Philosophical Magazine}, 12(120):1179--1192, 1965.

\bibitem{onofri2017full}
C~Onofri, M~Legros, J~L{\'e}chelle, H~Palancher, C~Baumier, C~Bachelet, and
  C~Sabathier.
\newblock Full characterization of dislocations in ion-irradiated
  polycrystalline uo2.
\newblock {\em Journal of Nuclear Materials}, 494:252--259, 2017.

\bibitem{bawane2021tem}
Kaustubh Bawane, Xiang Liu, Tiankai Yao, Marat Khafizov, Aaron French,
  J~Matthew Mann, Lin Shao, Jian Gan, David~H Hurley, and Lingfeng He.
\newblock Tem characterization of dislocation loops in proton irradiated single
  crystal tho2.
\newblock {\em Journal of Nuclear Materials}, 552:152998, 2021.

\end{thebibliography}
\end{document}